\documentclass{pasj00}
\draft

\begin{document}
\SetRunningHead{Y. Shioya et al.}{SDF}
\Received{yyyy/mm/dd}
\Accepted{yyyy/mm/dd}

\title{
New High-Redshift Galaxies at $z$ = 5.8 - 6.5
in the Subaru Deep Field
}

\author{Yasuhiro \textsc{Shioya} \altaffilmark{1}, 
Yoshiaki \textsc{Taniguchi} \altaffilmark{1}, 
Masaru \textsc{Ajiki} \altaffilmark{1}, 
Tohru \textsc{Nagao} \altaffilmark{1,2},\\
Takashi \textsc{Murayama} \altaffilmark{1}, 
Shunji \textsc{Sasaki} \altaffilmark{1}, 
Ryoko \textsc{Sumiya} \altaffilmark{1}, \\
Yuichiro \textsc{Hatakeyama} \altaffilmark{1},  \&
Nobunari \textsc{Kashikawa} \altaffilmark{3}

}

\email{shioya@astr.tohoku.ac.jp}

\altaffiltext{1}{Astronomical Institute, Graduate School of Science, Tohoku University, \\
                 Aramaki, Aoba, Sendai 980-8578}
\altaffiltext{2}{INAF --- Osservatorio Astrofisico di Arcetri,\\
                 Largo Enrico Fermi 5, 50125 Firenze, Italy}
\altaffiltext{3}{National Astronomical Observatory of Japan,
          2-21-1 Osawa, Mitaka, Tokyo 181-8588}

\KeyWords{galaxies : formation -- galaxies : evolution}

\maketitle

\begin{abstract}
 
In order to search for high-redshift galaxies around $z \sim 6$
in the Subaru Deep Field, we have investigated NB816-dropout
galaxies where NB816 is the narrowband filter centered at 
815nm with FWHM of 12.5 nm for the Suprime-Cam on the Subaru
Telescope. Since the NB816 imaging is so deep we can 
detect 10 well-defined NB816-dropout galaxies that we identify to lie 
at $z$ = 5.8 -- 6.5.  We discuss their observational properties.
 
\end{abstract}

\section{Introduction}

Surveys for high-redshift galaxies provide observational bases to
explore the formation and evolution of galaxies. A number of such
surveys have been made in this decade. 
Indeed, the optical broad-band color selection technique has been utilized to
find large samples of high-$z$ galaxies at $z \sim 3$ -- 7; i.e., 
Lyman break galaxies (LBGs)
(e.g., Lanzetta et al. 1996; Madau et al. 1996; Steidel et al. 1999;
Ouchi et al. 2004; Giavalisco et al. 2004; Stanway et al. 2004;
Bouwens et al. 2004). Deep optical surveys with a
narrowband filter have also been conducted to search for strong
Lyman $\alpha$ emitters (LAEs) at high redshift.
Since the early success of narrowband filter surveys (e.g., Hu et al.
1996; Cowie \& Hu 1998), a large number of high-$z$  LAEs have been
found  (e.g., Hu et al. 2002, 2004; Kodaira et al. 2003; 
Malhotra \& Rhoads 2001; Taniguchi et al. 2005; see for reviews,
Taniguchi et al. 2003; Spinrad 2004). Although such deep narrowband 
imaging data are basically used to find strong emission-line objects, 
they can also be used as one of the photometric data points. 
For example, Cuby et al. (2003) made their optical imaging survey for
LAEs at $z \sim 6.6$ by using both broad-band data ($B$, $V$, $R$, and $I$)
and narrow-band data taken with their NB920 filter. They found a LAE
at $z = 6.17$ from a sample of NB920-excess objects. In this case, 
the NB920 data were used to sample the continuum above the Lyman break.
This successful result suggested that any narrowband data can be used
to find not only strong LAEs but also so-called dropout high-$z$ galaxies.

Recently, Shioya et al. (2005) proposed a new method, intermediate-band dropout method, 
to search for high-$z$ galaxies (see also Fujita 2003).
In this method, an intermediate-band filter at $\lambda >$ 7000 \AA~ is used
together with typical $i^\prime$ and $z^\prime$ filters. 
We focus on dropouts using the intermediate-band filter, while Cuby et. 
al. (2003) selected a z=6.17 Lyman break galaxy using a narrowband filter.
The merit of the new method is to distinguish either very late-type stars such as
L and T dwarfs or dusty galaxies at intermediate redshift from real high-$z$ LBGs.
The reason for this is that such interlopers do not show strong narrow-band
depression  although they have very red colors that are indicative of LBGs.
In the survey method, we may use a narrowband filter at $\lambda >$ 7000 \AA ~
(Hu et al. 2004). 
We also note that Y.Kakazu (private communication) proposed the NB816-
depressed method to separate high-reshift quasars and galaxies from L and T
dwarfs based on the same principle.
In order to find high-$z$ galaxies around $z \sim 6$, we use
the data archive of the Subaru Deep Field (Kashikawa et al. 2004). Since 
in the SDF, deep imaging data with the narrowband filter NB816 are available, 
they are very useful to demonstrate that the narrowband dropout technique is
powerful to find such high-$z$ galaxies.

\section{The Narrow-Band Dropout Method}

Our new method is basically to find  narrow-band-dropout objects from a deep imaging
survey data set. In order to describe our method practically, we consider the case of
NB816 dropouts where NB816 is a narrow-band filter centered at $\lambda_{\rm c} \sim
815$ nm with a FWHM $=$ 12 nm (e.g., Taniguchi et al. 2003b and references therein; Hu et al. 2004).

In this case, we need deep imaging data taken with filters of $i^\prime$, $z^\prime$,
and NB816. In Fig. 1, we show the transmission curve of the three filters taken from
the data for Suprime-Cam filters\footnote{
http://www.subarutelescope.org/Observing/Instruments/SCam/sensitivity.html}
(Miyazaki et al. 2002).

We also show spectral energy distributions (SEDs) of a LBG (starburst) 
at $z \sim 5.9$, an L dwarf star in our Galaxy, and a dusty galaxy at $z \sim 1.1$. 
The SED of a LBG is produced by GALAXEV (Bruzual \& Charlot 2003). 
The SEDs of local galaxies are well reproduces by models whose star-formation 
rate declines exponentially (the $\tau$ model): i.e., $SFR(t) \propto \exp(-t/\tau)$, 
where $t$ is the age of galaxy and $\tau$ is the time scale of star formation. 
We set $\tau=1$ Gyr and $t=1$ Gyr to derive the SED of a starburst (SB) galaxy. 
To make an SED of an SB with emission lines, we calculate emission-line fluxes
using the formula of $L({\rm H}\beta)({\rm ergs \; s^{-1}})=4.76 \times 10^{-13}N_{\rm Lyc}$, 
where $N_{\rm Lyc}$ is the ionizing photon production rate (Leitherer \& Heckman 1995), 
and the table of relative luminosity to H$\beta$ luminosity in PEGASE 
(Fioc \& Rocca-Volmerange 1997). 
The SED of an LBG is the SB model with $A_V=0$ and 
that of a dusty galaxy is the SB model with $A_V=10$ mag adopting the reddening curve 
of Calzetti et al. (2000). 
It is shown that the LBG at $z=6$ is selected as a NB816 dropout.
However, the L dwarf is not selected as a NB816 dropout even though it shows a very
red  $i^\prime - z^\prime$  color. In this way, we can select LBGs at
$z>5.8$ as NB816 dropouts.

In Fig. 2, we show the diagram between  $i^\prime - z^\prime$ and $NB816 - z^\prime$
for galaxies found in the Subaru Deep Field (Kashikawa et al. 2004). 
The selection procedure is given in the next section.
Galactic stars (Gunn \& Stryker 1983) and L and T dwarfs\footnote{
http://spider.ipac.caltech.edu/staff/davy/ARCHIVE/} 
(Kirkpatrick 2003) are also shown.
An evolutionary path of a LBG is shown by the blue curve; 
its SED is generated by using the GALAXEV (Bruzual \& Charlot 2003).

When we select LBG candidates at $z \sim6$, we adopt a color criterion of
$i^\prime - z^\prime >$ 1.3 -- 1.5 (e.g., Nagao et al. 2004; Stanway et al. 2004;
Giavalisco et al. 2004). However, this criterion is not enough to select such LBGs
unambiguously because L and T dwarfs also have such very red colors.
In addition, some dusty galaxies at intermediate redshift may also have 
such red colors (e.g., a part of extremely red objects; see Hu \& Ridgway 1994). 

Here we would like to stress that LBGs at $z>5.8$ are efficiently selected 
if we also use another criterion of $NB816 - z >$ 1.5 -- 2; 
i.e., NB816-depressed objects.
The most important merit of this method is that LBG candidates are selected 
more unambiguously because this new criterion can separate LBGs from interlopers
(i.e., L, T dwarfs and dusty galaxies at lower redshift).
Hu et al. (2004) noted that LAEs at $z \simeq 5.7$ are well separated from 
L and T dwarfs if a proper criterion on the $NB816 - z^\prime$ color is adopted. 

\section{Application to the Subaru Deep Field Data}

In order to demonstrate how the new NB dropout method works well,
we apply it to the Subaru Deep Field data (Kashikawa et al. 2004).
Photometric catalogs (Version 1) have been made public\footnote{ 
http://soaps.naoj.org/sdf/}. 
In the following analysis, we use a $2^{\prime \prime} \phi$ aperture magnitude 
and the error in $2^{\prime \prime} \phi$ aperture magnitudes from the catalog. 
We also calculate errors of colors using the 
errors in the $2^{\prime \prime} \phi$ aperture magnitude. 

As shown in Fig.2, being independent of the strength of the Lyman 
$\alpha$ emission, galaxies at $z > 5.73$ can be identified as 
NB816-dropout objects. 
Assuming that there are no elliptical-like red galaxies at $z \sim 2$, 
we estimate a reddest $NB816 - z^\prime$ color of $\approx$ 1.3,
corresponding to the color of elliptical galaxies at $z \sim 1$. 
Therefore, we adopt the following selection criterion for
NB816-dropout galaxies,
\begin{equation}
NB816 - z^\prime > 1.5
\end{equation}
and 
\begin{equation}
z^\prime < 26.07 \; (5 \sigma). 
\end{equation}

In order to reduce any contaminations from foreground objects 
that are free from absorption by the intergalactic neutral hydrogen, 
we also adopt following criteria
\begin{eqnarray}
B & > & 28.45 \; (3 \sigma), \\
V & > & 27.74 \; (3 \sigma), \\
{\rm and} \; \; \; R & > & 27.80 (3 \sigma). 
\end{eqnarray}
By using the above three criteria, we first obtain a sample of 42 NB816-dropout objects. 
From 42 NB816-dropout objects, 
we select the objects with $NB816 - z^\prime > 1.5$ above the $3 \sigma$ error 
as the more reliable NB-dropout objects. 
The number of more reliable NB816-dropout objects is 15. 

Next, we must separate high-$z$ galaxies from Galactic low-temperature stars 
(L and T dwarfs). It is known that 
most L and T dwarfs satisfy the following relation: 

\begin{equation}
(i^\prime - z^\prime) - (NB816 - z^\prime) \simeq 0.8, 
\end{equation}
as shown in Fig. 2.
Taking account of the scatter of the colors of L and T dwarfs, 
we can separate LBG candidates from them by using the following criterion,

\begin{equation}
(i^\prime - z^\prime) - (NB816 - z^\prime) < 0.5.
\end{equation}
In order to securely select a reliable sample of NB816 dropout galaxies,
we pick up objects that satisfy the condition (7) above 
the $3 \sigma$ error conditions from 15 more reliable NB816-dropout objects. 
Then we finally obtain a sample of 10 NB816-dropout galaxies. 
Their photometric properties are summarized in Table 1. 
In this table, we present $3 \sigma$ error of colors. 

We note that galaxies with faint continuum and strong-emission lines in $z^\prime$, 
e.g., strong [OIII] emission-line objects at $z \sim 0.8$, may satisfy 
the above selection criteria. 
Actually, Y. Kakazu has found strong [OIII] emission-line objects 
in her NB816-depressed sample (private communication).
Since there is an NB921 observation for SDF, 
we can separate such objects from genuine NB816 dropouts using $z^\prime - NB921$ colors. 
If strong-emission lines lie in both $z^\prime$ and $NB921$, 
those galaxies would be observed as $NB921$-excess objects. 
If strong-emission lines do not lie in $NB921$ but in $z^\prime$, 
$NB921$-band of those galaxies must be very faint. 
All our {\it NB816}-dropout sample are detected ($> 3 \sigma$) in {\it NB921}-band, and 
their $z^\prime - NB921$ colors are nearly flat, which range 
from $-0.39$ to $0.02$. 
Therefore, they are considered to be galaxies at redshift between
5.73 and 6.50.
One object, ID 35247, is considered to be a LAE 
since its $z^\prime - NB921$ color is 1.49. 
In fact, this object is identified as SDF J132408.3+271543 in Taniguchi et al. (2005) 
and confirmed as a LAE at $z = 6.554$. 

Now, we consider the UV luminosity function of star-forming galaxies at $z \sim 6$. 
In addition to the UV luminosity functions of star-forming galaxies at $z \sim 3$
to $\sim 5$  (e.g., Steidel et al. 1999; Ouchi et al. 2004), 
those of star-forming galaxies at $z \sim 6$ were also recently 
obtained by red $i^\prime - z^\prime$ color ($i^\prime$-dropout galaxies, 
e.g., Bouwens et al. 2004) or by using the $i^\prime - z_R$ versus $z_B - z_R$ 
diagram (Shimasaku et al. 2004). Hu et al. (2004) also obtained the UV luminosity 
function of Lyman $\alpha$ emitters (LAEs) at $z \sim 5.7$. 

Since the observed $z^\prime$-magnitude of a galaxy depends on 
not only the UV luminosity but also the redshift of the galaxy and 
the strength of the Ly$\alpha$ emission line, 
we compare our results with the prediction evaluated under some assumptions. 
If the UV luminosity function of galaxies at $z \sim 6$ is 
the same as that at $z \sim 4.0$ or 4.7 (Ouchi et al. 2004), 
we can evaluate the number of star-forming galaxies at $z \sim 5.8$ to 6.6. 
We show the expected galaxy number evaluated using the SB model SED without 
emission lines in Fig.3. 
We also show the expected number of galaxies for 
the UV luminosity function of LBGs at $z \sim 6$ (Bouwens et al. 2004) and 
that of LAEs at $z \sim 5.7$ (Hu et al. 2004). 
Since the strong Ly$\alpha$ emission line modify the $z^\prime$ magnitude  
for galaxies at redshift higher than $\approx$ 5.75, 
we use our SB model with emission lines for the UV luminosity function of LAEs. 
The rest-frame equivalent width of Ly$\alpha$ emission is about 70 \AA~ 
in our SB model with emission lines. 
Taking account of the completeness for the SDF sample as a function of $z^\prime$ 
magnitude (Kashikawa et al. 2004), we expect that the number of galaxies brighter
than $z^\prime = 26.0$ ranges from 42 to 200 (Fig.3). 

The selection criteria of our final sample are so severe that 
the number of our final sample (10) may give a lower limit. 
On the other hand, the number of our first sample (42) may give a upper limit, 
since it may include contaminations. 
Fig.3 suggests that the number of LBGs at $z \sim 6$ is 
smaller than that at $z \sim 4$ -- 5. 
This trend is consistent with that derived from the observation of 
the Hubble Ultra Deep Field (Bunker et al. 2004) and 
the Hubble Ultra Deep parallel fields (UDF PFs: Bouwens et al. 2004). 
Although the reason why the number of our sample is smaller than 
the prediction using the UV luminosity function at $z \sim 6$ is still not clear, 
we note that the number density of bright LBGs at $z \sim 6$ in the SDF 
is 60 \% of the value obtained in the UDF PFs by Bouwens et al. (2004) 
(Shimasaku et al. 2005). 

\section{REMARKS}

In Section 3, we have demonstrated that the NB816 dropout method is capable of
selecting reliable candidates of LBGs at $z>5.8$. Although we have adopted
the NB816 dropout method there, our method is also useful when we use other
narrow-band filters.

For example, NB921 is useful to find galaxies at $z > 6.7$ while 
NB711 is useful for galaxies at $z > 4.9$.
It is also noted that a rest-frame H$\alpha$ filter (NB656)
and a rest-frame [O {\sc iii}] filter can be used to search for
galaxies at $z > 4.4$ and at $z > 3.1$, respectively.
Since there are a number of deep survey data in which some of narrow-band filters
are used, it is recommended that the narrow-band dropout method is applied
to them, providing new samples of LBGs/LAEs at high redshift.


We would like to thank to the member of the Subaru Deep Field project. 
We would like to thank the referee, Esther Hu, for her useful
comments and suggestions.
We also thank Yuko Kakazu for providing us valuable information 
prior to the publication.
This work was financially supported in part by
the Ministry of Education, Culture, Sports, Science, and Technology
(Nos. 10044052, and 10304013) and JSPS (No. 15340059).
MA and TN are JSPS fellows.

\clearpage

\begin{figure}
\begin{center}
\FigureFile(80mm,80mm){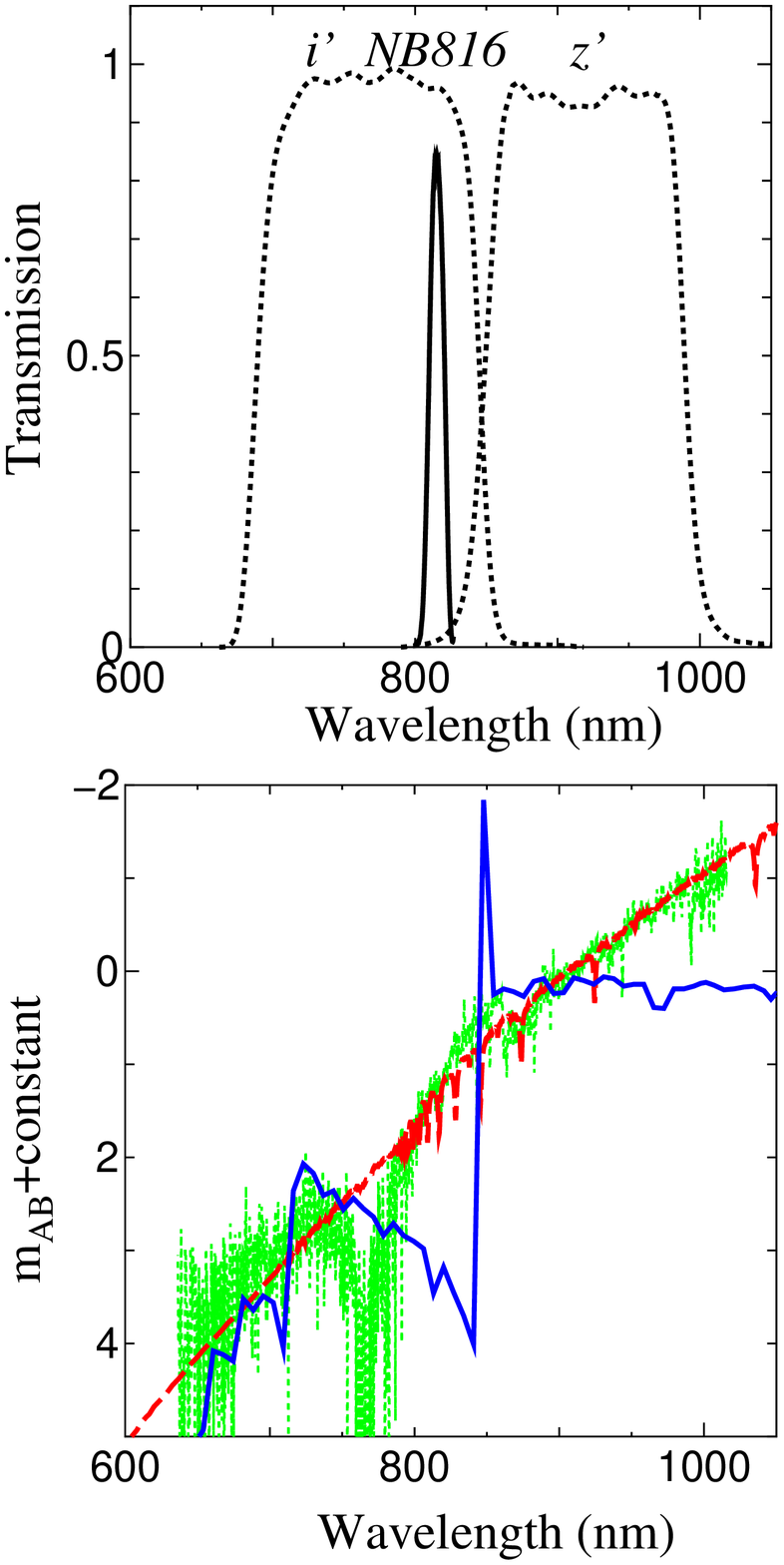}
\end{center}
\caption{
The upper panel shows the transmission curves of the filters, 
$i^\prime$, $z^\prime$, and NB816. 
The lower panel shows SED of 
Lyman break galaxy (LBG) at $z \sim 5.9$ (blue), 
dusty starburst at $z \sim 1.1$ (red), 
and Galactic L dwarf stars, 2MASS J09510549+3558021 (green). 
The SEDs of LBG and dusty starburst are produced by GALAXEV 
(Bruzual, Charlot 2003, see text). 
Although, these objects shows the same $i^\prime - z^\prime$ colors ($\sim 2$), 
The $NB816-z^\prime$ colors of LBG is redder than those of 
a L dwarf star and a dusty starburst galaxy (see Fig.2).
}\label{fig:fig1}
\end{figure}

\begin{figure}
\begin{center}
\FigureFile(80mm,80mm){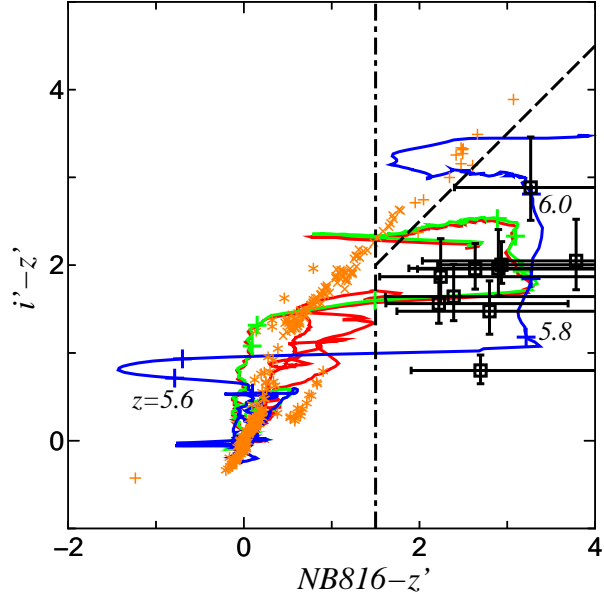}
\end{center}
\caption{
$i^\prime - z^\prime$ vs. $NB8168-z^\prime$ diagram displaying 
the colors of model galaxies and stars. 
Solid lines are loci of model galaxies produced by GALAXEV (Bruzual, Charlot 2003): 
red line - $\tau = 1$ Gyr model with age of 8 Gyr, 
green line - $\tau = 1$ Gyr model with age of 1 Gyr (SB model), 
and blue line - SB model with emission lines (see text). 
Orange marks show stars. Asterisks are Gunn \& Stryker (1983)'s star, 
crosses are L dwarfs and pluses are T dwarfs.
Open squares with error bars show our final sample.
}\label{fig:fig2}
\end{figure}

\begin{figure}
\begin{center}
\FigureFile(80mm,80mm){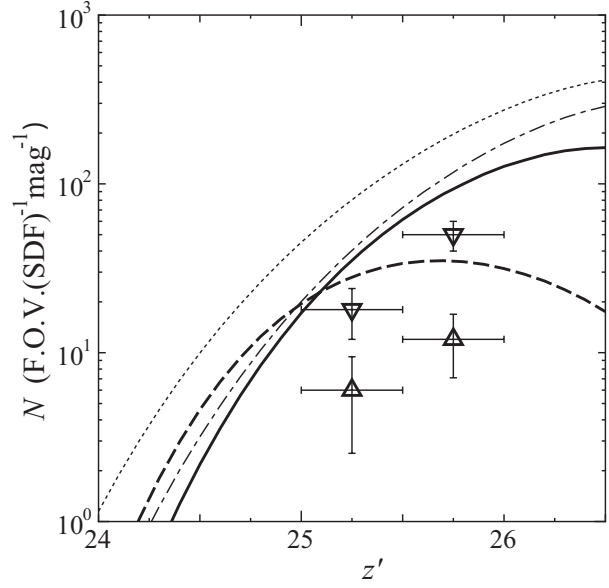}
\end{center}
\caption{
Number of NB816-dropout galaxies as a function of $z^\prime$ magnitude. 
Triangles show the number of our final sample and 
inverse triangles show the number of first sample. 
Lines show the expected number of galaxies with $z > 5.73$ for different assumptions. 
Thin dotted (dot-dashed) line shows the case for the SED without Ly$\alpha$ emission 
together with the UV luminosity function at $z \sim 4.0$ (4.7) (Ouchi et al. 2004). 
Thick solid line shows the case for the SED without Ly$\alpha$ emission 
together with the UV luminosity function at $z \sim 6.0$ (Bouwens et al. 2004) and 
thick dashed line shows the case for the SED with Ly$\alpha$ emission 
together with the UV luminosity function at $z \sim 5.7$ (Hu et al. 2004).
}\label{fig:fig3}
\end{figure}

\clearpage

\begin{table}
\begin{center}
\caption{Photometric properties of NB816-dropout galaxies}
\begin{tabular}{cccccccc}
\hline
\hline
  ID    & Name                  & \multicolumn{4}{c}{Optical AB magnitude} &  \multicolumn{2}{c}{Optical colors}                  \\
        &                       & $i^\prime$ & $z^\prime$ & NB816 & NB921 & $i^\prime - z^\prime$     & $NB816 - z^\prime$ \\
\hline
  18309 &  SDF J132436.6+271246 &  27.93  & 25.89 &  29.67 &  26.01 &   $2.05_{-0.33}^{+0.47}$&   $3.78_{-1.75}^{+\infty}$ \\
  35247 &  SDF J132408.3+271543 &  27.62  & 25.98 &  28.37 &  24.49 &   $1.64_{-0.27}^{+0.37}$&   $2.39_{-0.78}^{+\infty}$ \\
  39694 &  SDF J132418.4+271633 &  27.33  & 25.77 &  27.99 &  25.75 &   $1.56_{-0.22}^{+0.28}$&   $2.22_{-0.60}^{+1.47}  $ \\
  46427 &  SDF J132444.8+271749 &  27.82  & 25.95 &  28.20 &  25.96 &   $1.87_{-0.31}^{+0.43}$&   $2.24_{-0.69}^{+2.45}  $ \\
  66699 &  SDF J132406.0+272222 &  27.86  & 25.90 &  28.79 &  26.00 &   $1.96_{-0.31}^{+0.44}$&   $2.90_{-1.02}^{+\infty}$ \\
  77541 &  SDF J132344.7+272450 &  27.52  & 26.05 &  28.84 &  26.06 &   $1.48_{-0.26}^{+0.35}$&   $2.80_{-1.05}^{+\infty}$ \\
  94464 &  SDF J132452.1+272842 &  27.34  & 25.34 &  28.27 &  25.23 &   $2.02_{-0.21}^{+0.26}$&   $2.93_{-0.73}^{+3.29}  $ \\
 106928 &  SDF J132339.7+273138 &  27.44  & 25.49 &  28.12 &  25.66 &   $1.96_{-0.23}^{+0.29}$&   $2.63_{-0.66}^{+1.94}  $ \\
 117034 &  SDF J132436.0+273356 &  26.50  & 25.70 &  28.40 &  25.93 &   $0.80_{-0.15}^{+0.18}$&   $2.70_{-0.79}^{+\infty}$ \\
 176181 &  SDF J132507.7+274509 &  28.16  & 25.28 &  28.54 &  25.49 &   $2.89_{-0.37}^{+0.56}$&   $3.27_{-0.87}^{+\infty}$ \\
\hline
\end{tabular}
\end{center}
\end{table}


\begin{thebibliography}{}

\bibitem[Ajiki et al. (2002)]{ajiki02}
Ajiki, M., et al. 2002, \apj, 576, L25

\bibitem[Ajiki et al. (2003)]{ajiki03}
Ajiki, M., et al. 2003, AJ, 126, 2091

\bibitem[Ajiki et al. (2004)]{ajiki04}
Ajiki, M., et al. 2004, PASJ, 56, 597

\bibitem[Bouwens et al. (2003)]{bouwes03}
Bouwens, R. J., et al. 2003, \apj, 595, 589

\bibitem[Bouwens et al. (2004)]{bouwes04}
Bouwens, R. J., et al. 2004, \apj, 616, L79

\bibitem[Bruzual \& Charlot (2003)]{bruzual03}
Bruzual, G., \& Charlot, S. 2003, \mnras, 344, 1000

\bibitem[Bunker et al. (2004)]{bunker04}
Bunker, Andrew J., Stanway, Elizabeth R., Ellis, Richard S., \& McMahon, Richard G. 
2004, \mnras,355, 374

\bibitem[Calzetti et al. (2000)]{calzetti00}
Calzetti, D., Armus, L., Bohlin, R. C., Kinney, A. L., Koornneef, J., \& Storchi-Bergmann, T. 
2000, \apj, 533, 682

\bibitem[Cowie \& Hu (1998)]{cowie98}
Cowie, L. L., \& Hu, E. M. 1998, AJ, 115,1319

\bibitem[Cuby et al. (2003)]{cuby03}
Cuby, J.-G., Le Fevre, O., McCracken, H., Cuillandre, J.-C., Magnier, E., Meneux, B. 
2003, \aap, 405, L19

\bibitem[Dey et al. (1998)]{dey98}
Dey, A., Spinrad, H., Stern, D., Graham, J. R., \& Chaffee, F. H. 
1998, ApJ, 498, L93

\bibitem[Dickinson et al. (2004)]{dickinson04}
Dickinson, M., et al. 2004, \apj, 600, L99

\bibitem[Fioc \& Rocca-Volmerange (1997)]{fioc97}
Fioc, M., \& Rocca-Volmerange, B. 1997, \aa, 326, 950

\bibitem[Furusawa et al. (2000)]{furusawa00}
Furusawa, H., Shimasaku, K., Doi, M., \& Okamura, S. \apj, 534, 624

\bibitem[Giavalisco et al. (2004)]{giavalisco04}
Giavalisco, M., Dickinson, M., Ferguson, H. C., Ravindranath, C., 
Kretchmer, C., Moustakas, L. A., Madau, P., Fall, S. M., et al. 
2004, ApJ, 600, L103

\bibitem[Gunn \& Stryker (1983)]{gunn83}
Gunn, J. E., \& Stryker, L. L. 1983, \apjs, 52, 121

\bibitem[Hogg et al. (1998)]{hogg98}
Hogg, D. W., et al. 1998, \aj, 115, 1418

\bibitem[Hu \& Ridgway (1994)]{hu94}
Hu, E. M., \& Ridgway, S. E. 1994, AJ, 107, 1303

\bibitem[Hu et al. (2003)]{hu03}
Hu, E. M., McMahon, R. G., \& Egami, E. 1996, \apj, 459, L53

\bibitem[Hu et al. (2002)]{hu02}
Hu, E. M., Cowie, L. L., McMahon, R. G., Capak, P., Iwamuro, F., 
Kneib, J. -P., Maihara, T., \& Motohara, K. 2002, ApJ, 568, L75

\bibitem[Hu et al. (2004)]{hu04}
Hu, E. M., Cowie, L. L., Capak, P., McMahon, R. G., Hayashino, T., \& 
Komiyama, Y. 2004, AJ, 127, 563

\bibitem[Kashikawa et al. (2004)]{kashikawa04}
Kashikawa, N., et al. 2004, PASJ, 56, 1011

\bibitem[Kirkpatrick (2003)]{kirkpatrick03}
Kirkpatrick, J. Davy 2003, 
Brown Dwarfs, Proceedings of IAU Symposium No.211, (Ed.) Eduardo Martin. San Francisco, 
Astronomical Society of the Pacific, p. 189

\bibitem[Kodaira et al. (2003)]{kodaira03}
Kodaira, K., et al. 2003, PASJ, 55, L17

\bibitem[Lanzetta et al. (1996)]{lanzetta96}Lanzetta, K. M., 
Yahil, A., \& Fern\'{a}ndez-Soto, A. 1996, Nature, 381, 759

\bibitem[Leitherer \& Heckman (1995)]{leithere95}
Leitherer, C., \& Heckman, T. M. 1995, \apjs, 96, 9

\bibitem[Loeb \& Barkana (2001)]{loeb01}
Loeb, A., \& Barkana, R. 2001, ARA\&A, 39, 19

\bibitem[Madau et al. (1996)]{madau96}
Madau, P., Ferguson, H. C., Dickinson, M. E., Giavalisco, M., 
Steidel, C. C., \& Fruchter, A. 1996, MNRAS, 283, 1388

\bibitem[Miyazaki et al. (2002)]{miyazaki02}
Miyazaki, S., et al. 2002, PASJ, 54, 833

\bibitem[Nagao et al. (2004)]{nagao04}
Nagao, T., et al. 2004, \apj, 613, L9

\bibitem[Ouchi (2004)]{ouchi04}
Ouchi, M., Shimasaku, K., Okamura, S., Furusawa, H., Kashikawa, N., 
Ota, K., Doi, M., Hamabe, M., et al. 2004, ApJ, 611, 660

\bibitem[Partridge \& Peebles (1967)]{partridge67}
Partridge, R. B., \& Peebles, P. J. E. 1967, ApJ, 147, 868

\bibitem[Shimasaku et al. (2005)]{shimasaku05}
Shimasaku, K., Ouchi, M., Furusawa, H., Yoshida, M., Kashikawa, N., Okamura, S 
2005, PASJ, in press (astro-ph/0504373)

\bibitem[Shioya et al. (2005)]{shioya05}
Shioya, Y., Taniguchi, Y., Ajiki, M., Nagao, T., Murayama, T., Sasaki, S., Sumiya, R., 
Hatakeyama, Y. 2005, PASJ, 57, 287

\bibitem[Spinrad (2004)]{spinrad04}
Spinrad, H. 2004, in Astrophysics Update, J.W.Mason, J. (ed.),  
(Berlin: Springer-Verlag and Chichester, UK: Praxis Publishing), p.155

\bibitem[Stanway et al. (2004)]{stanway04}
Stanway, E. R., Bunker, Andrew J., McMahon, R. G., Ellis, R. S., 
Treu, T., McCarthy, P. J. 2004, \apj, 607, 704

\bibitem[Steidel \& Hamilton (1992)]{steidel92}
Steidel, C. C., Hamilton, D. 1992, AJ, 104, 941

\bibitem[Steidel et al. (1999)]{steidel99}
Steidel, Charles C., Adelberger, Kurt L., Giavalisco, Mauro, Dickinson, Mark, \& Pettini, Max 1999, \apj, 519, 1

\bibitem[Taniguchi et al. (2003a)]{taniguchi03a}
Taniguchi, Y. et al. 2003a, ApJ, 585, L97

\bibitem[Taniguchi et al. (2003b)]{taniguchi03b}
Taniguchi, Y. et al. 2003b, JKAS, 36, 123 (erratum 36, 283)

\bibitem[Taniguchi et al. (2005)]{taniguchi05}
Taniguchi, Y., et al. 2005, PASJ, 57, 165

\end{thebibliography}
\end{document}